\documentclass[sigconf]{acmart}
\renewcommand\footnotetextcopyrightpermission[1]{}  
\usepackage{multirow}
\usepackage{fancyvrb} 
\usepackage{graphicx}  
\usepackage{subcaption} 
\usepackage{listings}  
\usepackage{xcolor}    
\usepackage{tabularx}
\usepackage[utf8]{inputenc}  
\usepackage{pifont}
\usepackage[utf8]{inputenc} 
\usepackage{listings}
\usepackage{listingsutf8} 

\usepackage{newunicodechar}  

\usepackage{amssymb}         

\usepackage[T1]{fontenc}
\usepackage[utf8]{inputenc}
\usepackage{CJKutf8}  

\lstdefinestyle{custom}{
  backgroundcolor=\color{white},
  basicstyle=\ttfamily\small,
  breaklines=true,
  numbers=left,
  numberstyle=\tiny\color{gray},
  keywordstyle=\color{blue},
  commentstyle=\color{green!50!black},
  stringstyle=\color{orange},
  showstringspaces=false
}

\lstdefinelanguage{JavaScript}{
  keywords={break, case, catch, class, continue, const, debugger, default, delete, do, else, export, extends, finally, for, function, if, import, in, instanceof, let, new, return, super, switch, this, throw, try, typeof, var, void, while, with, yield, await, async},
  sensitive=true,
  morecomment=[l]//,
  morecomment=[s]{/*}{*/},
  morestring=[b]",
  morestring=[b]'
}

\lstdefinelanguage{JavaScript}{
  keywords={function,if,else,return,var,let,const,for,while,switch,case,break,continue,new,class,this,super,import,export,default,async,await,try,catch,finally},
  keywordstyle=\color{blue}\bfseries,
  ndkeywords={true,false,null,undefined},
  ndkeywordstyle=\color{magenta},
  identifierstyle=\color{black},
  sensitive=true,
  comment=[l]{//},
  morecomment=[s]{/*}{*/},
  stringstyle=\color{red},
  morestring=[b]',
  morestring=[b]"
}

\lstdefinelanguage{CSS}{
  keywords={
    width, height, border-style, border-width, border-color, 
    clip-path, filter, drop-shadow, polygon, rgba, transparent
  },
  keywordstyle=\color{blue}\bfseries,
  ndkeywords={.plane-body},  
  ndkeywordstyle=\color{purple}\bfseries,
  sensitive=true,
  morecomment=[s]{/*}{*/},  
  morestring=[b]',
  morestring=[b]",
  alsodigit={-}
}

\lstdefinelanguage{HTML5}{
  language=HTML,
  sensitive=true, 
  alsoletter={<>=-},
  morecomment=[s]{<!--}{-->},
  morestring=[b]",
  morestring=[b]',
  morekeywords={
    html, head, body, title, meta, link, script, div, span, 
    h1, h2, h3, h4, h5, h6, p, a, img, ul, ol, li, table, 
    tr, td, th, form, input, button, select, option, textarea
  },
  morekeywords=[2]{
    color, background-color, font-size, font-family, margin, 
    padding, border, width, height, display, position, top, 
    left, right, bottom,border-style 
  },
  morekeywords=[3]{
    function, var, let, const, if, else, for, while, return, 
    document, window, alert, console, log, querySelector
  },
  keywordstyle=\color{blue},
  keywordstyle=[2]{\color{purple}},
  keywordstyle=[3]{\color{orange}},
  commentstyle=\color{green!50!black},
  stringstyle=\color{red}
}

\lstdefinestyle{custom}{
  breaklines=true,         
  breakatwhitespace=false, 
  basicstyle=\footnotesize\ttfamily, 
  keywordstyle=\color{blue}, 
  commentstyle=\color{green!50!black}, 
  stringstyle=\color{red},  
  numbers=left,            
  numberstyle=\small\color{gray}, 
  frame=single,            
  showspaces=false,        
  showstringspaces=false,  
}

\AtBeginDocument{%
  }


\begin{document}

\title{FrontendBench: A Benchmark for Evaluating LLMs on Front-End Development via Automatic Evaluation}

\author{Hongda Zhu}
\authornote{Equal contribution}
\affiliation{
  \institution{ByteDance}
  \city{Chengdu}
  \country{China}
}
\email{zhuhongda@bytedance.com}

\author{Yiwen Zhang}
\authornotemark[1]
\affiliation{
  \institution{ByteDance}
  \city{Beijing}
  \country{China}
}
\email{yiwenzhangvincent@163.com}

\author{Bing Zhao}
\authornote{Corresponding author}
\affiliation{
  \institution{ByteDance}
  \city{Beijing}
  \country{China}
}
\email{zhaobingcars@gmail.com}

\author{Jingzhe Ding}
\affiliation{
  \institution{ByteDance}
  \city{Beijing}
  \country{China}
}
\email{dingjingzhe@bytedance.com}

\author{Siyao Liu}
\affiliation{
  \institution{ByteDance}
  \city{Beijing}
  \country{China}
}
\email{liusiyao.sine@bytedance.com}

\author{Tong Liu}
\affiliation{
  \institution{ByteDance}
  \city{Chengdu}
  \country{China}
}
\email{liutong.chestnut@bytedance.com}

\author{Dandan Wang}
\affiliation{
  \institution{ByteDance}
  \city{Beijing}
  \country{China}
}
\email{wdd846009505@gmail.com}

\author{Yanan Liu}
\affiliation{
  \institution{ByteDance}
  \city{Beijing}
  \country{China}
}
\email{liuyanan.1211@jiyunhudong.com}

\author{Zhaojian Li}
\affiliation{
  \institution{ByteDance}
  \city{Shanghai}
  \country{China}
}
\email{lizhaojian.joeli@bytedance.com}
\begin{abstract}
Large Language Models (LLMs) have made significant strides in front-end code generation. However, existing benchmarks exhibit several critical limitations: many tasks are overly simplistic, test cases often lack rigor, and end-to-end validation is absent. These issues hinder the accurate assessment of model performance. To address these challenges, we present FrontendBench, a benchmark co-developed by humans and LLMs. FrontendBench categorizes tasks based on code functionality and incorporates interactive test scenarios, enabling a more comprehensive and practical evaluation of front-end code generation capabilities. The benchmark comprises 148 meticulously crafted prompt–test case pairs spanning five levels of web components, from basic UI elements to complex interactive features. Each task reflects realistic front-end development challenges. Furthermore, we introduce an automatic evaluation framework that executes generated code within a sandbox environment and assesses outcomes using predefined test scripts. This framework achieves a 90.54\% agreement rate with expert human evaluations, demonstrating high reliability. We benchmark several state-of-the-art LLMs on FrontendBench and observe substantial performance disparities in handling real-world front-end tasks. These results highlight FrontendBench as a reliable and scalable benchmark, supporting consistent multimodal evaluation and providing a robust foundation for future research in front-end code generation. Our data and code will be released soon.
\end{abstract}





\maketitle

\section{Introduction}

Large Language Models (LLMs) could generate executable front-end code based on natural-language queries, significantly improving productivity and meeting the demands of efficient software development \cite{zhu2023minigpt4}. A variety of mainstream code evaluation benchmarks in recent have contributed to the advancement of code generation research \cite{iyer-etal-2018-mapping,lu2021codexglue,takamoto2023pdebench,jain2024,chai2024mcevalmassivelymultilingualcode,song2025}. However, the rapid advancement of LLMs and code generation systems based on them has exposed the growing inadequacies of existing benchmarks. Specifically, the capabilities of LLMs in handling more complex and challenging real-world scenarios remain insufficiently evaluated. In particular, current benchmarks fall short in effectively capturing the quality and interactivity required for front-end code generation tasks, which typically involve rich user interfaces, dynamic interactions, and close integration between visual elements and underlying logic. Tab. ~\ref{tab:compar} compares general-purpose and front-end-specific benchmarks in terms of task difficulty, unit testing, and end-to-end comprehension. 

\begin{table*}

  \small
  \begin{tabular}{p{0.25\textwidth}p{0.35\textwidth}p{0.35\textwidth}}
    \toprule
    \multirow{2}{*}{\textbf{Evaluation Dimension}} & 
    \multicolumn{2}{c}{\textbf{Benchmark Type}} \\
    \cmidrule(lr){2-3}
    & \textbf{General Code} & \textbf{Front-End Code} \\
    \midrule
    \textbf{Task Difficulty} & 
    \begin{itemize}
        \item Covers basic programming tasks with simple logic and static scenarios
        \item High model scores reduce the benchmark's discriminative power
    \end{itemize} & 
    \begin{itemize}
        \item Includes complex tasks requiring both functional correctness and visual presentation accuracy
    \end{itemize} \\
    \midrule
    \textbf{Unit Testing} & 
    \begin{itemize}
        \item Low-quality test cases with incomplete reference outputs
        \item Missing ground truths introduce evaluation bias
        \item Redundant samples affect accuracy
        \item Front-end code coverage is limited
    \end{itemize} & 
    \begin{itemize}
        \item Focused on real-world front-end tasks
        \item More complete and representative test coverage
        \item Targeted assessment of UI/UX behaviors
    \end{itemize} \\
    \midrule
    \textbf{End-to-End Comprehensiveness} & 
    \begin{itemize}
        \item Evaluates code logic and syntax, but lacks user interaction modeling
    \end{itemize} & 
    \begin{itemize}
        \item Captures full user interaction flow through end-to-end validation
    \end{itemize} \\
    \bottomrule
  \end{tabular}
  
    \caption{Comparison between general benchmarks and front-end benchmarks.}
     \label{tab:compar}
\end{table*}

\begin{itemize}

\item {\texttt{Task Difficulty:}} General code benchmarks primarily target basic programming tasks \cite{chen2021evaluating}, whereas front-end benchmarks involve more complex and interactive scenarios.

\item {\texttt{Unit Testing:}} The data quality of many general code benchmarks is low. As a result, they fail to provide effective evaluation for front-end code generation. \cite{hendrycks2021measuring,manh2025codemmlu,li2022competition}. In contrast, front-end benchmarks provide more targeted and reliable evaluation.

\item {\texttt{End-to-End Comprehensiveness:}} General code benchmarks emphasize logic correctness and adherence to coding standards \cite{tong2024codejudge,jain2025testgenevalrealworldunit}, while front-end benchmarks focus more on user interactivity and dynamic behavior.

\end{itemize}

  Therefore, a more specialized and higher-quality benchmark is required to address challenges in task complexity, unit testing, and end-to-end evaluation for front-end code generation. To address persistent challenges in task complexity, unit testing, and end-to-end evaluation for front-end code generation, we introduce \textbf{FrontendBench}, a specialized and high-quality benchmark designed for the rigorous assessment of LLMs in front-end programming scenarios. FrontendBench emphasizes interactive behavior and organizes a total of 148 carefully curated tasks across five application categories: concept explanation, utilities, games, web interfaces, and data visualizations. Each task is constructed to reflect realistic development challenges and varying levels of complexity.

To enable scalable and reliable evaluation, we develop an automatic evaluation framework that executes the generated code within a sandboxed environment using task-specific test scripts. These scripts are iteratively refined based on pilot experiments and expert feedback to ensure high precision and coverage, thereby ensuring their effectiveness and robustness. Empirical evaluation demonstrates that the framework achieves a 90.54\% agreement rate with expert human judgments.

Furthermore, we evaluate a range of state-of-the-art LLMs on FrontendBench to characterize their capabilities and limitations in handling diverse front-end tasks. The results highlight notable performance disparities among models, underscoring the need for continued advancements in model reasoning and interaction understanding in the front-end domain.

Our main contributions are as follows:
\begin{enumerate}
    \item We construct the FrontendBench dataset, which categorizes web components into multiple application types and covers a wide range of complex front-end scenarios.
    \item We propose an automatic evaluation framework that verifies element presence, functional correctness, and interactive logic, addressing both visual and end-to-end interaction challenges in real-world scenarios.
    \item We adopt a dual-verification mechanism that combines sandbox execution and manual inspection to ensure evaluation accuracy.
\end{enumerate}

This benchmark, along with its evaluation framework, serves as a valuable resource for understanding the front-end development capabilities of contemporary LLMs. By offering fine-grained assessments of model performance across diverse interactive coding tasks, FrontendBench also provides critical insights into current model strengths and weaknesses. Moreover, it offers practical guidance for the iterative improvement of AI-powered code assistance tools, facilitating their advancement toward more reliable and context-aware front-end development support.

\section{Related Work}

\subsection{General Code Benchmark}

In recent years, the use of LLMs in automatic code generation has grown rapidly. As a result, code quality benchmarking has become an increasingly important research topic.Starting from typical benchmarks like HumanEval~\cite{chen2021evaluating}  and MBPP~\cite{austin2021program}, increasingly benchmarks are constructed to evaluate the coding capabilities of LLMs.

Li et al. proposes EvoCodeBench. This benchmark collects data from multiple real-world code repositories. It aligns with practical development settings by considering factors such as code distribution and dependency structures \cite{li2024evocode}. They also introduced the first repository-level code generation task. This task improves the benchmark’s relevance to real-world software engineering. Agashe et al. focuses on the Jupyter Notebook environment. They proposed JuICe, which is the first large-scale dataset for interactive code generation \cite{agashe-etal-2019-juice}. However, EvoCodeBench and JuICe are limited to Python and specific task types, which restricts their generalizability across languages and applications. Zhu et al. propose DOMAINEVAL to improve evaluation coverage. This benchmark includes six domains, such as computation, networking, and cryptography \cite{zhu2024}. However, it also focuses only on Python. In addition, the data distribution across domains is unbalanced. To support multilingual evaluation, Peng et al. propose HumanEval-XL \cite{peng2024humaneval-xl}. It includes 23 natural languages and 12 programming languages. This design enables a broader evaluation of LLMs' ability to generalize across languages.

Although these benchmarks show strong generalization in algorithmic code generation, they do not address front-end code evaluation well. Most of them focus on logic and structure, but they fail to capture the interactive and visual nature of front-end code.

\subsection{Front-end Code Benchmark}

Recent studies have explored various evaluation indicators for front-end code generation. Most of these works focus on image-to-code generation tasks and evaluate model performance primarily through similarity-based metrics. Several representative datasets for evaluating front-end code generation are summarized below:

MultiPL-E is the first multilingual code generation benchmark. It covers a wide range of programming paradigms and includes a small subset of front-end code, thereby offering foundational support for front-end code generation tasks \cite{10103177}. However, its scope is not tailored to front-end code, and the dataset only contains a limited amount of front-end-related content. Consequently, it does not capture the unique requirements and complexity of front-end code generation.
Laurençon et al.~\cite{laurenccon2024unlocking} propose WebSight, a large-scale synthetic dataset consisting of 2 million HTML–screenshot pairs, aimed at supporting front-end code generation. While WebSight facilitates image-to-code learning, its evaluation primarily focuses on code-level similarity, potentially encouraging models to learn superficial patterns rather than capturing deeper visual semantics essential to front-end design. 
To better capture visual characteristics of front-end design, Si et al. propose the Design2Code benchmark, which consists of webpage screenshot-code pairs and evaluates performance based on visual similarity between generated pages and reference images \cite{si2024design2code}. However, it focuses solely on visual appearance and omits considerations of end-to-end interactivity, and thus exhibits limited generalizability across broader front-end tasks. 
Furthermore, Yun et al. introduce the Web2Code benchmark to address the issue of low-quality front-end code datasets \cite{yun2024web2code}. However, the evaluation criterion is limited to visual similarity between the generated and reference webpage screenshots. This highlights the narrow focus of the evaluation, which lacks a rating of interactive functionalities.

Although previous studies provide useful benchmarks, they still have clear limitations in evaluating front-end code. Some of them ignore end-to-end interaction, while others only consider visual similarity or lack diversity in evaluation tasks. Building on prior work, our benchmark offers a more comprehensive and practical solution by evaluating not only front-end code generation tasks but also their behavior in real user interactions, enhancing both generalization and real-world relevance

\section{FrontendBench}

\subsection{Dataset Construction}

To accomplish a comprehensive and automatic evaluation of LLMs in front-end development, we aim at constructing a benchmark that not only provides detailed task descriptions, but also includes corresponding test scripts designed to reliably assess the correctness and completeness of any code generated for a given task, regardless of its implementation. As illustrated in Fig.~\ref{fig:dataset_construction}, the dataset construction process follows four main steps:

\begin{figure*}
      \centering
      \includegraphics[width=1\linewidth]{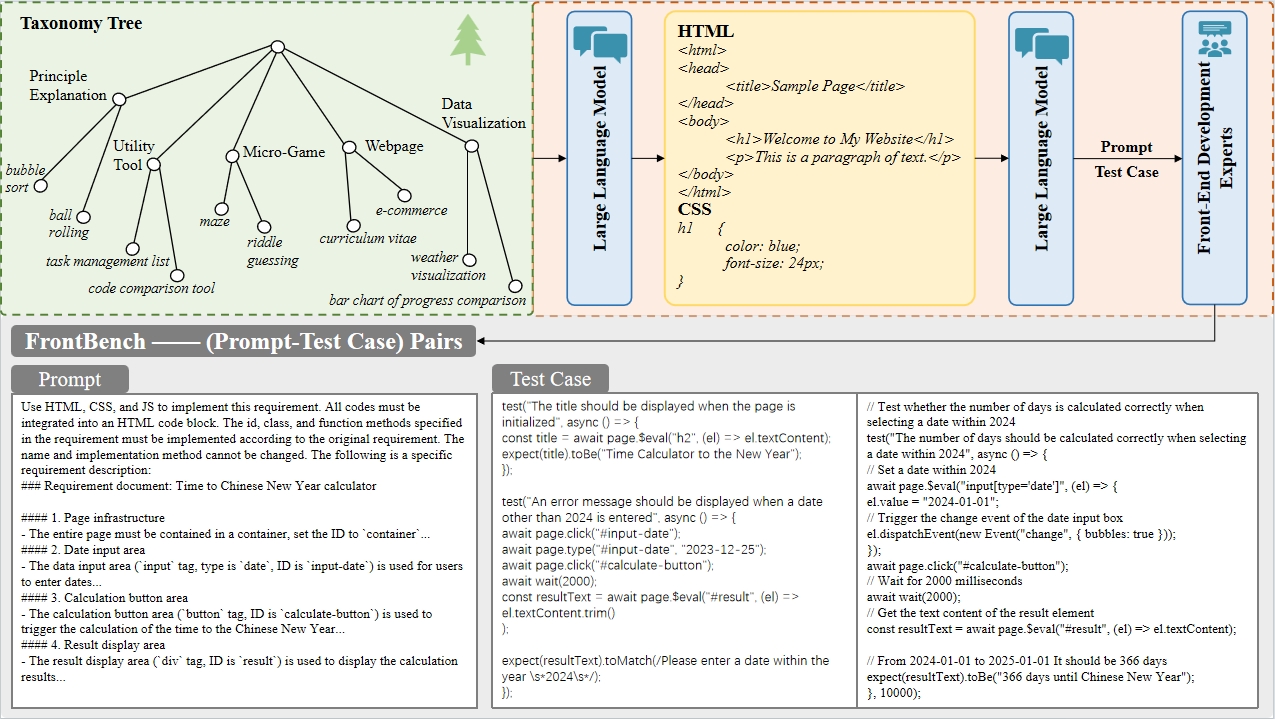}
   
      \label{fig:enter-label}

    \caption{Construction process and examples of FrontendBench.}
  \Description{Construction Process and Examples of FrontendBench}
  \label{fig:dataset_construction}
\end{figure*}

\begin{enumerate}
    \item \textbf{Application Categorization:} Based on a survey of common front-end use cases, we define five categories: demonstrations, tools, mini-games, webpages, and data visualizations.

    \item \textbf{Instance Specification:} For each category, we design multiple application instances (e.g., “bubble sort” for demonstrations, “to-do list” for tools), each paired with a concise functional description.

    \item \textbf{LLM-Driven Generation:} For each instance, an LLM is prompted to generate HTML code. This output is then used to reverse-engineer a detailed prompt and corresponding test code, ensuring consistency between specification and verification. Key DOM elements are explicitly identified (e.g., via \texttt{id}, \texttt{class}) and styled with fixed visual attributes.

    \item \textbf{Post-processing and Packaging:} The generated test code is parsed to extract executable units, which are integrated into standardized test templates using Jest and Puppeteer. The result is a structured dataset consisting of paired \texttt{prompt} and \texttt{test\_code} entries.
\end{enumerate}

\subsection{Dataset Optimization}

The raw dataset constructed in Section 3.1 heavily depends on the LLM’s intrinsic code analysis capabilities. However, it overlooks certain boundary conditions, such as the determination of color values and the specification of DOM element dimensions. Moreover, the issues revealed by the HTML outputs of different models are often inconsistent. Therefore, further refinement of the dataset is necessary to improve its reliability and evaluation consistency.

To facilitate efficient optimization, we adopt an iterative human-in-the-loop strategy, in which human evaluation follows a defined scoring rubric: If all functionalities specified in the prompt are correctly implemented, no functional errors are observed, and the rendered page appears visually sound upon manual inspection, a score of 1 is assigned; otherwise, the score is 0. All human experts involved in the evaluation have at least 1 year of software development experience.

The iterative process consists of the following steps:
\begin{enumerate}
    \item Run experiments on FrontendBench and collect the automatic evaluation scores for each model.
    \item Organize a human review team to render the model-generated HTML locally and score each output based on the human evaluation rubric.
    \item Domain experts analyze discrepancies between human and machine scores, and resolve issues such as ambiguous prompt descriptions or misaligned test case assertions.
    \item The revised data set is reintroduced into the evaluation pipeline and the above steps are repeated iteratively until a high agreement is achieved between the human and automatic evaluation results.
\end{enumerate}

\subsection{FrontendBench Dataset Composition}

Finally, we construct our benchmark dataset, \textbf{FrontendBench}, which comprises 148 test cases, each consisting of a paired \texttt{prompt} and corresponding \texttt{test\_code}. Each \texttt{prompt} delineates the desired functionality and page layout, incorporating specific visual styles and explicit DOM identifiers (e.g., \texttt{id}, \texttt{class}) to support automatic evaluation. The associated \texttt{test\_code} integrates a reusable testing template with scripted user interactions and assertions, implemented using Puppeteer and Jest. To enable precise analysis during the evaluation phase, we further classify the benchmark into five difficulty levels, as illustrated in Tab.~\ref{tab:difficulty_levels}.

\begin{table}[h]
\centering
\begin{tabularx}{\linewidth}{ccc}
\toprule
\textbf{Level} & \textbf{Description} & \textbf{Number} \\
\midrule
1 & Simple static page & 9 \\
2 & Simple page with dynamic effects & 18 \\
3 & Simple page with basic interaction & 90 \\
4 & Simple page with complex interaction & 22 \\
5 & Complex page with complex interaction & 9 \\
\bottomrule
\end{tabularx}
\caption{Difficulty levels and question distribution in FrontendBench.}
\label{tab:difficulty_levels}
\end{table}

\section{Automatic Evaluation Framework for FrontendBench}

\begin{figure*}[h]
  \centering
  \includegraphics[width=0.7\textwidth]{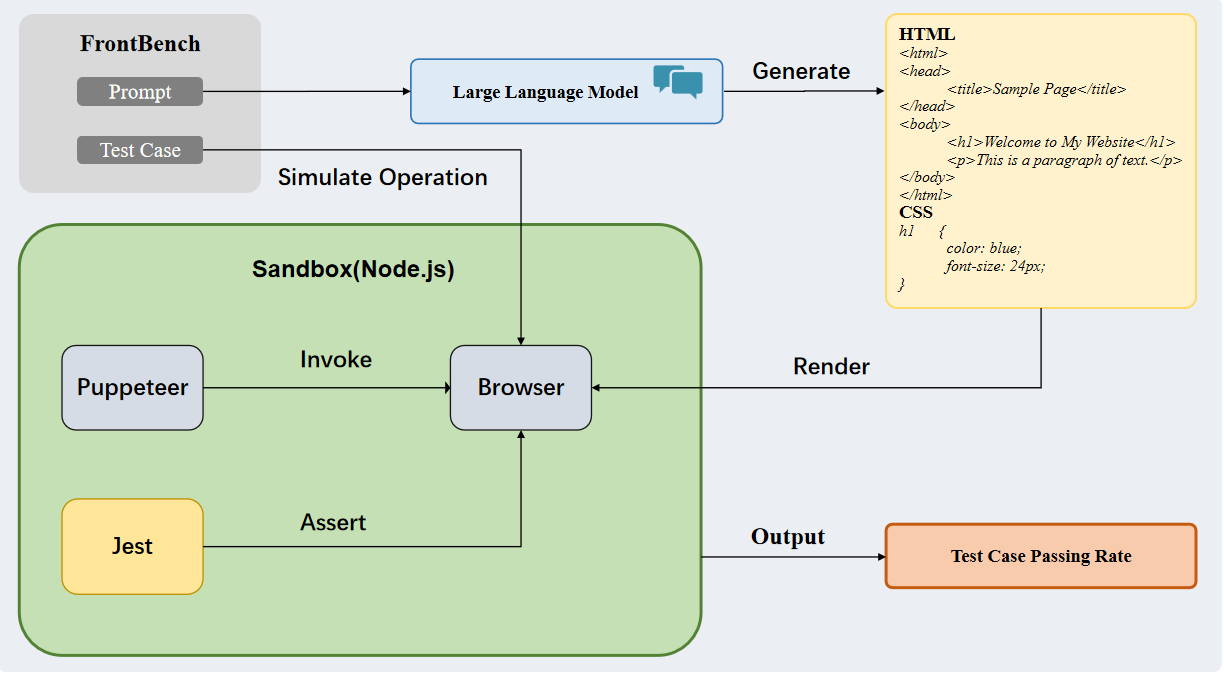}
  \caption{Front-end code generation eutomatic evaluation method.}
  \Description{Front-end code generation automatic evaluation method}
  \label{fig:evaluation_method}
\end{figure*}

We establish an isolated sandbox environment equipped with all necessary external dependencies, including Node.js, Puppeteer, and Jest. Node.js serves as the runtime environment for executing JavaScript; Puppeteer provides headless browser instances to simulate user interactions; and Jest offers assertion APIs for validating test outcomes. After each interaction is simulated, the success or failure of the test is determined based on the resulting page state.

To conduct the evaluation, the constructed dataset (FrontendBench) is loaded into the sandbox. The HTML code generated by the model under evaluation is provided as input, along with its corresponding test script. Within this setup, Puppeteer APIs are used to launch and control the browser, execute predefined user operations, and verify whether the generated HTML satisfies the required functional specifications.

Given the varying complexity of tasks, each evaluation item contains a different number of test cases. To ensure fairness and emphasize full-task correctness, we adopt a binary scoring strategy: a score of 1 is awarded only if all test cases for a given evaluation item pass; otherwise, the score is 0. Thus, each evaluation item contributes either 0 or 1 point to the model’s overall score.

As illustrated in Fig.~\ref{fig:evaluation_method}, the evaluation process proceeds through the following steps:

\begin{enumerate}
    \item A sandbox environment is prepared with pre-installed dependencies, including Node.js, Puppeteer, and Jest. Multiple LLMs are selected as evaluation targets.

    \item Using the experimental platform, the FrontendBench dataset is loaded in parallel using multi-threading. For each evaluation item, the \texttt{prompt} field is passed as input to the model under evaluation. Once the model generates its HTML output, the sandbox is invoked, and both the model-generated HTML and the corresponding \texttt{test\_code} are loaded into the environment.

    \item Within the sandbox, the test cases are executed as follows:
    \begin{itemize}
        \item The sandbox reads and runs the test script associated with the current evaluation item.
        \item Puppeteer launches a headless browser instance for page rendering.
        \item User interactions (e.g., clicks, scrolling) are simulated via Puppeteer APIs.
        \item The resulting page state is captured through Puppeteer.
        \item Jest assertions evaluate the outcome by comparing actual results against expected conditions.
    \end{itemize}

    \item Scoring and reporting are performed as follows:
    \begin{itemize}
        \item Each evaluation item is assigned a score based on the success or failure of its test cases.
        \item Scores are accumulated across all items to compute the overall model performance.
        \item Statistical reports are generated to summarize and visualize the evaluation results.

    \end{itemize}
\end{enumerate}

\section{Experiment}

\subsection{Experimental Settings}

To ensure a comprehensive and representative evaluation, we selected a diverse set of LLMs, covering both close-source and open-source systems. The evaluated models include Gemini-2.5-pro~\cite{team2023gemini}, DeepSeek-R1~\cite{guo2025deepseek}, DeepSeek-V3~\cite{liu2024deepseek}, and OpenAI-o3-mini~\cite{openai2025o3o4mini}.

The evaluation process consists of two components: automatic evaluation and human evaluation.  We independently calculate pass rate to facilitate stable comparisons of model performance across iterations. The pass rate is calculated as follows, where \(\mathit{model\_score}\) denotes the total automatic evaluation score of a model, and \(\mathit{total\_score}\) represents the maximum possible score across the dataset:

\begin{equation}
\mathit{PassRate} = \frac{\mathit{model\_score}}{\mathit{total\_score}} \times 100\%
\end{equation}

Besides model performance, we also pay attention to the consistency within 2 evaluation methods. For each evaluation item, if the auto-eval score differs from the human-eval score, the discrepancy counter (\(\mathit{diff\_count}\)) is incremented by one. Based on this, we calculate the human-machine consistency rate for the evaluation results. Reflecting data reliability ,the consistency rate is calculated using the following formula, where \(\mathit{diff\_count}\) denotes the number of mismatches and \(\mathit{total\_count}\) refers to the total number of evaluation items:

\begin{equation}
\mathit{ConsistencyRate} = \left(1 - \frac{\mathit{diff\_count}}{\mathit{total\_count}}\right) \times 100\%
\end{equation}

\subsection{Experiment Results and Analysis}

\begin{table*}[t]

  \centering
  \begin{tabular}{lcccc} 
    \toprule
    \textbf{Metric} & \textbf{Gemini-2.5-pro} & \textbf{DeepSeek-R1}& \textbf{DeepSeek-V3}& \textbf{o3-mini} \\
    \midrule
    \textbf{Auto-Eval Pass Rate} & 70.27& 75.00& 66.89& 83.11\\
    \textbf{Human-Eval Pass Rate} & 79.73& 67.57& 68.24& 83.78\\
    \textbf{Consistency Rate (\%)} & 87.83& 89.86& 91.89& 92.57\\
    \bottomrule
  \end{tabular}
    \caption{Experimental results(\%) across different models.}
    \label{tab:benchmark_results}

\end{table*}

As shown in Tab.~\ref{tab:benchmark_results},we can find that o3-mini achieves the highest pass rate at 83.11\%, followed by DeepSeek-R1 (75.00\%), Gemini-2.5-pro (70.27\%), and DeepSeek-V3 (66.89\%).  However, the human evaluation pass rates for Gemini-2.5-pro and DeepSeek-R1 are 7–9 percentage points higher than their corresponding automatic evaluation scores. This result reflects the differences in the capabilities of different models in front - end tasks.
Moreover, the human-machine consistency rate for all evaluated models exceeds 87\%, with an average of approximately 90.54\%. The small variance among consistency rates suggests that the automatic evaluation results are generally reliable. Based on this, we proceed with an overall analysis of the automatic evaluation results. 

For each difficulty level, we also calculate the consistency rate between human and automatic evaluations for each model. The diff rate is calculated as follows, where \(\mathit{model\_diff\_count}\) denotes the number of discrepancies for a given model at the current difficulty level, and \(\mathit{level\_total\_count}\) is the total number of evaluation items at that level:

\begin{equation}
\mathit{LevelConsistentRate} =\left( 1- \frac{\mathit{model\_diff\_count}}{\mathit{level\_total\_count}} \right) \times 100\%
\end{equation}

\begin{table*}[t]

  \centering
  \begin{tabular}{ccccc}
    \toprule
    \textbf{Level}& \textbf{Gemini-2.5-pro} & \textbf{DeepSeek-R1}& \textbf{DeepSeek-V3}& \textbf{o3-mini} \\
    \midrule
    1& 55.56& 88.89& 100.00& 100.00\\
    2& 94.44& 100.00& 88.89& 88.89\\
    3& 91.11& 86.67& 90.00& 93.33\\
    4& 90.91& 90.91& 100.00& 86.36\\
    5& 66.67& 100.00& 88.89& 100.00\\
    \bottomrule
  \end{tabular}
    \caption{Consistency rates(\%) of each model across difficulty levels.}
    \label{tab:consistencystats}
\end{table*}

As shown in Tab.~\ref{tab:consistencystats}, the agreement between automatic evaluation and human judgment varies across different task difficulty levels, as reflected by the diff rates (complementary to consistency rates). While DeepSeek-R1 and o3-mini maintain high consistency even at higher difficulty levels, Gemini-2.5-pro exhibit greater variability, particularly under very simple or complex tasks. These results suggest that the reliability of automatic evaluation may be affected by task difficulty and model-specific generation behaviors. This highlights the importance of difficulty-aware evaluation and supports the use of consistency analysis as a means to monitor the robustness of our automatic assessment framework across diverse scenarios.

We also compute the automatic evaluation scores and corresponding pass rates for each model across different difficulty levels. The pass rate is calculated as follows, where \(\mathit{model\_pass\_score}\) denotes the score obtained by the model at the current difficulty level, and \(\mathit{level\_total\_score}\) represents the maximum achievable score at that level:

\begin{equation}
\mathit{LevelPassRate} = \frac{\mathit{model\_pass\_score}}{\mathit{level\_total\_score}} \times 100\%
\end{equation}

\begin{table*}[t]

  \centering
  \begin{tabular}{ccccc}
    \toprule
    \textbf{Level}& \textbf{Gemini-2.5-pro} & \textbf{DeepSeek-R1}& \textbf{DeepSeek-V3}& \textbf{o3-mini} \\
    \midrule
     1& 66.67 & 55.56 & 66.67 & 66.67 \\
    2& 72.22 & 66.67 & 66.67 & 77.78 \\
    3& 73.33 & 81.11 & 70.00 & 85.56 \\
    4& 63.64 & 68.18 & 59.09 & 81.82 \\
    5& 55.56 & 66.67 & 55.56 & 88.89 \\
    \bottomrule
  \end{tabular}
    \caption{Model pass rates(\%) in Auto-eval across difficulty levels.}
  \label{tab:llm_pass_rates}
\end{table*}

As shown in Tab.~\ref{tab:llm_pass_rates}, the pass rate of most models initially increase and then decline as task difficulty rises, suggesting that these models perform best on evaluation items of moderate difficulty but struggle with more complex tasks. In contrast, o3-mini exhibits a generally increasing trend across difficulty levels, indicating stronger capabilities in complicated front-end code generation tasks.

\subsection{Case study}

To analyze the consistency between automatic evaluation and human evaluation, we select 2 representative results.

\textbf{Case Type 1: The automatic evaluation passed and the human evaluation passed.}  
  
   \textbf{Key test case logic:}
\begin{lstlisting}[style=custom, language=JavaScript]
✓ Page initialization is successful (83 ms)
✓ Input box function works properly (83 ms)
✓ Send wish function works properly (125 ms)
✓ Paper airplane animation works properly (4601 ms)
✓ Test whether the paper airplane flies out of the slot (118 ms)
✓ Test the color of the paper airplane (116 ms)
✓ Test that the mailbox is at the bottom left (42 ms)
Test Suites: 1 passed, 1 total
Tests: 7 passed, 7 total
Snapshots: 0 total
Time: 7.359 s
Ran all test suites.
\end{lstlisting}

\begin{figure}
        \centering
        \includegraphics[width=0.5\linewidth]{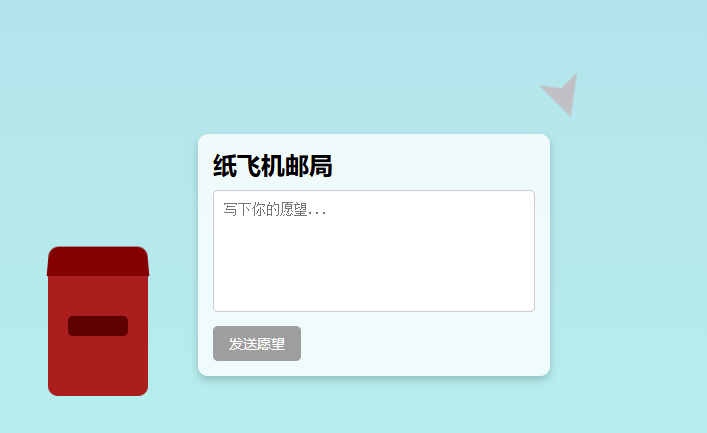}
        \caption{Rendered output of the paper airplane element.}
        \label{fig:success}
    \end{figure}
        
    \label{screen-1}

The rendered output is shown in Fig.~\ref{fig:success}. Based on the above information, all use cases of the automatic evaluation have passed, and the human evaluation has also passed, this indicates that there is a high degree of consistency between the two.

\textbf{Case Type 2: The automatic evaluation failed and the human evaluation failed.}  

  \textbf{Key test case logic:}
\begin{lstlisting}[style=custom,language=JavaScript]
✓  Page should load correctly (124 ms)
✓ English text should be encoded correctly (77 ms)
✓ English text should be decoded correctly (93 ms)
✕ Chinese text should be encoded correctly (107 ms)
✕ Chinese text should be decoded correctly (72 ms)
✓ Error should be displayed when decoding invalid Base64 (105 ms)
✓ All input and output should be cleared (81 ms)

*Test template > Chinese text should be encoded correctly
expect(received).toContain(expected)
Expected substring: "5L2g5aW9"
Received string: ""
\end{lstlisting}

\begin{figure}[htbp]
  \centering
  \begin{subfigure}[b]{0.23\textwidth}
    \includegraphics[width=\textwidth]{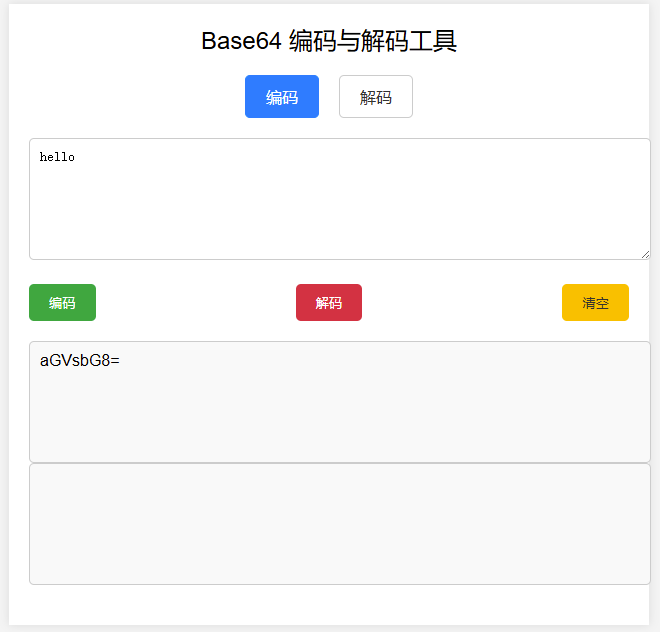}
    \caption{Test using English}
    \label{fig:sub1}
  \end{subfigure}
  \hfill
  \begin{subfigure}[b]{0.23\textwidth}
    \includegraphics[width=\textwidth]{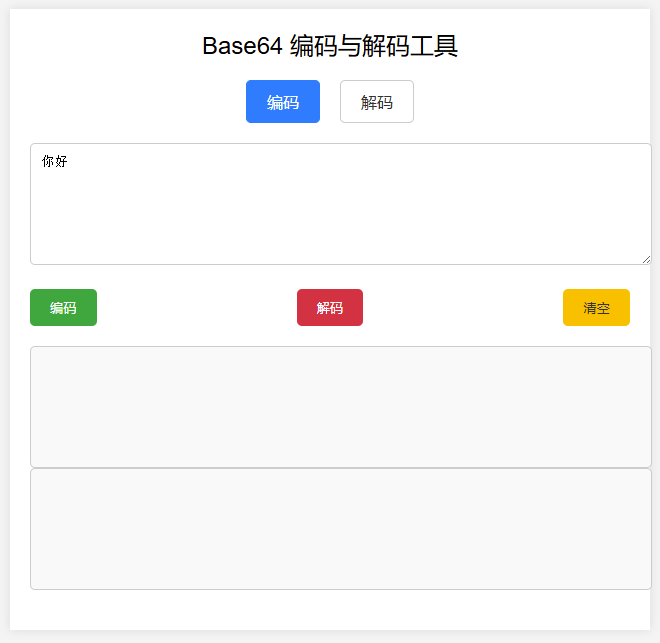}
    \caption{Test using Chinese}
    \label{fig:sub2}
  \end{subfigure}
  \caption{Encoding behavior with different input types.}
  \label{fig:dual}
\end{figure}

The rendered output is shown in Fig.~\ref{fig:dual}. Based on the above information, the prompt requires the text to be encoded as a Base64 string. The model uses \texttt{btoa} to perform encoding, which works for ASCII input such as English, as confirmed in Fig.~\ref{fig:sub1}. However, as shown in Fig.~\ref{fig:sub2}, it fails when encoding non-ASCII characters like Chinese, due to \texttt{btoa}'s limitation. Besides, the use cases for Chinese encoding and decoding in the automatic evaluation failed, This indicates that the automatic evaluation and the human evaluation are highly consistent.

\section{Conclusion}

We present FrontendBench, a high-quality benchmark that integrates fine-grained functional specifications with executable test cases to support realistic front-end development scenarios. Through human–LLM collaboration, the framework effectively reduces dataset noise and evaluation bias while maintaining broad coverage. Experimental results demonstrate strong consistency between automatic evaluation and expert human assessments, validating the framework’s reliability for scalable and objective performance measurement. This work not only strengthens the methodological foundations of front-end code generation research but also provides practical insights for optimizing LLM training and fine-tuning workflows.

\appendix

\section{Case: Financial Management Product Page}

\subsection{Requirement Description}

The case requires the development of an integrated financial product display single-page application, supporting multi-category filtering and dynamic product showcasing features. The system should be implemented using HTML, CSS, and JavaScript, employing a card-based grid layout, and be compatible with mainstream browsers and mobile devices.

\begin{enumerate}
    \item \textbf{Product Filtering Mechanism}
    \begin{itemize}
        \item A \texttt{Dropdown} menu will be used to filter five categories of products: All, Funds, Insurance, Trusts, and Wealth Management.
        \item By default, all products are displayed. Once a category is selected by the user, the displayed content will dynamically update.
    \end{itemize}

    \item \textbf{Product Display Design}
    \begin{itemize}
        \item A responsive grid layout will be adopted, with 2 to 4 columns per row on desktop devices and automatic adjustment to a single column on mobile devices.
        \item Unified card style with borders, rounded corners, and shadows will be used for product display.
        \item Each product card will contain the product name as the heading and detailed description information.
    \end{itemize}

\end{enumerate}

\subsection{Generation Code}

Taking the Gemini model as an example, the following is the front-end code automatically generated by the model based on the requirements, with the CSS and JS sections omitted.

\begin{lstlisting}[style=custom,language=HTML5]
<!DOCTYPE html>
<html lang="en">
<head>
    <meta charset="UTF-8">
    <meta name="viewport" content="width=device-width, initial-scale=1.0">
    <title>Financial Product Shelf</title>
    <style> ... </style>
</head>
<body>
    <div class="container" id="page-container-id">
        <h1 class="page-title" id="page-title-id">Financial Product Shelf</h1>
        <div class="filters" id="filters-id">
            <select class="category-select" id="category-select-id">
                <option value="all">All Products</option>
                <option value="fund">Fund</option>
                <option value="insurance">Insurance</option>
                <option value="trust">Trust</option>
                <option value="wealth">Wealth Management</option>
            </select>
        </div>

        <div class="product-grid" id="product-grid-id">
            <!-- Product cards will be dynamically inserted here by JavaScript -->
        </div>
    </div>

    <script> ... </script>

</body>
</html>
\end{lstlisting}

\subsection{Test Code}

The following is the evaluation code for the front-end code, with test cases for the interactive components as shown below:

\begin{lstlisting}[style=custom,language=JavaScript]
test('Should display all products when the page initializes', async () => {
  const productCards = await page.$$('.product-card');
  expect(productCards.length).toBe(8);
});

test('Filter function should work correctly - Select fund category', async () => {
  await page.select('#category-select-id', 'fund');
  const productCards = await page.$$('.product-card');
  expect(productCards.length).toBe(2);
});

test('Filter function should work correctly - Select all categories', async () => {
  await page.select('#category-select-id', 'all');
  const productCards = await page.$$('.product-card');
  expect(productCards.length).toBe(8);
});

test('Product cards should contain product name and description', async () => {
  const productCard = await page.$('.product-card');
  const productName = await productCard.$('.product-name');
  const productDescription = await productCard.$('.product-description');
  expect(productName).toBeTruthy();
  expect(productDescription).toBeTruthy();
});
\end{lstlisting}

\subsection{Test Result}

This test covers a total of 5 items, primarily aimed at verifying the relevant functionalities of the financial wealth management product display single-page application. The specific test items are as follows:

\begin{enumerate}
    \item The page should contain all necessary DOM elements.
    \item All products should be displayed upon page initialization.
    \item The function of selecting the "Funds" category should work correctly.
    \item The function of selecting the "All" category should work correctly.
    \item The product cards should include the product name and description.
\end{enumerate}

Taking the front-end code generated by the Gemini model as an example, all tests passed, with the following results:

\begin{itemize}
    \item The test suite execution time was 2.648 seconds.
    \item All 5 tests passed, with no failed cases.
\end{itemize}

\begin{lstlisting}[style=custom]
PASS ./tmpwzmjmjyx.test.ts (10.48 s)
  Test Template
    Main container and basic elements exist (pass, 330 ms)
    Undo and Redo buttons are initially disabled (pass, 83 ms)
    Can draw on the canvas (pass, 1123 ms)
    Undo button becomes available after drawing and can reverse actions (pass, 2897 ms)
    Redo button becomes available after undoing and can restore actions (pass, 3164 ms)
    Line width controller updates displayed value correctly (pass, 52 ms)
    Color picker modifies color properly (pass, 93 ms)
    \end{lstlisting}
\begin{lstlisting}[style=custom]
Test Suites: 1 passed, 1 total
Tests:       7 passed, 7 total
Snapshots:   0 total
Time:        10.521 s
Ran all test suites.
\end{lstlisting}

The test results indicate that the wealth management product display page generated by the Gemini model meets the expected functionality, with stable core interaction logic and performance that meets basic usage requirements.

\begin{figure}[htbp]
  \centering
  \includegraphics[width=0.4\textwidth]{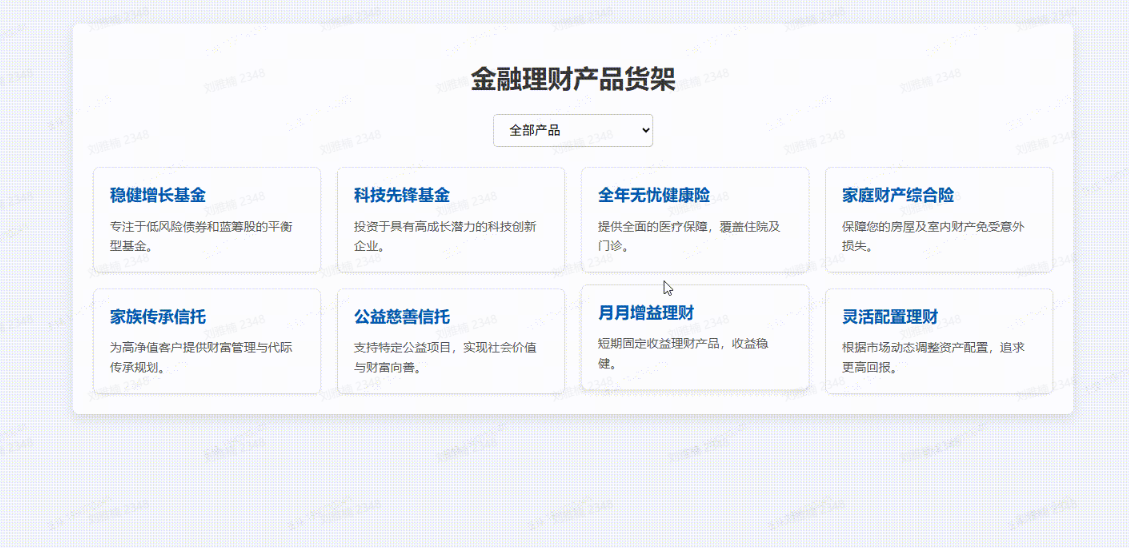}
  \caption{The screenshot of the financial wealth management product}
  \label{screen-1}
\end{figure}


\bibliographystyle{ACM-Reference-Format}
\bibliography{bib}

\end{document}